\documentclass[10pt]{article}
\usepackage[OE]{express}

\begin{document}

\title{Resonant absorption of electromagnetic waves in transition anisotropic media}

\author{Kihong Kim\authormark{*}}

\address{Department of Energy Systems
Research and Department of Physics, Ajou University, Suwon 16499, Korea}

\email{\authormark{*}khkim@ajou.ac.kr}

\begin{abstract}
We study the mode conversion and resonant absorption phenomena occurring in a slab of a stratified
anisotropic medium, optical axes of which are tilted with respect to the direction of inhomogeneity, using the invariant imbedding
theory of wave propagation.
When the tilt angle is zero, mode conversion occurs if the longitudinal component of the permittivity tensor, which is the one
in the direction of inhomogeneity in the non-tilted case,
varies from positive to negative values within the medium, while the transverse component plays no role.
When the tilt angle is nonzero, the wave transmission and absorption show an asymmetry under the sign change of the incident
angle in a range of the tilt angle, while the reflection is always symmetric. We calculate the reflectance,
the transmittance and the absorptance for several configurations of the permittivity tensor and find that resonant absorption is
greatly enhanced when the medium from the incident surface to the resonance region is hyperbolic than when
it is elliptic. For certain configurations, the transmittance and absorptance curves display sharp peaks at some incident angles
determined by the tilt angle.
\end{abstract}


\section{Introduction}

In recent years, there have been many studies on the strong absorption of electromagnetic (EM) waves occurring in media with near-zero effective
dielectric permittivity $\epsilon$, termed epsilon-near-zero media, and in transition metamaterials inside which $\epsilon$ varies continuously
from positive to negative values, passing through a narrow resonance region where the real part of $\epsilon$ is identically zero [1--14].
In the presence of an infinitesimally small amount of damping in the resonance region, a large absorption of the EM wave energy occurs.
This strong resonant absorption arises due to
the mode conversion of incident $p$-polarized EM waves into longitudinal plasma oscillations at the resonance region [1, 5, 15--23]. In this paper, we study theoretically similar phenomena occurring in transition anisotropic media with unequal values
of the transverse and longitudinal components of the permittivity tensor, using the invariant imbedding theory of wave propagation.

The characteristics of EM wave propagation in anisotropic media, such as uniaxial and biaxial media, have been studied extensively
for a long time. More recently, hyperbolic media, where one of the principal values of the permittivity tensor has a different sign from the
others and their dispersion surface is a hyperboloid, have attracted great interest from many researchers [24--26].
The strong absorption of EM waves in uniform anisotropic media has been studied recently by several authors, with an emphasis
on the role of a hyperbolic dispersion relation [9--14].
In this paper, we consider an inhomogeneous slab of a stratified
anisotropic medium, optical axes of which are tilted with respect to the direction of inhomogeneity.
We are interested in the case where the tensor components in the longitudinal and/or transverse
directions vary from positive to negative values within the medium and the associated mode conversion and resonant absorption phenomena.
We examine the different roles of the transverse and longitudinal components on resonant absorption.
We also investigate interesting asymmetric transmission and absorption phenomena occurring when the tilt angle is nonzero.

\section{Model}

We consider an anisotropic dielectric medium, the dielectric permittivity tensor of which is written as
\begin{eqnarray}
\epsilon^\prime=\left(\begin{array}{ccc}\epsilon_1 & 0 & 0\\
0 & \epsilon_2 &0\\ 0 & 0 & \epsilon_3\\\end{array}\right)
\end{eqnarray}
in the coordinate system $(x^\prime,y^\prime,z^\prime)$.
We assume that the primed coordinate system is obtained by rotating the unprimed coordinate system $(x,y,z)$ by an angle $\phi$
with respect to the $y$ axis, as depicted in Fig.~1. Then the permittivity tensor $\epsilon$ in the unprimed coordinate system is
related to $\epsilon^\prime$ by
\begin{equation}
\epsilon=R^T\epsilon^\prime R,
\end{equation}
where $R$ is the rotation matrix given by
\begin{equation}
R=\left(\begin{array}{ccc}\cos\phi & 0 & -\sin\phi \\ 0 & 1 &0 \\ \sin\phi & 0 & \cos\phi\\\end{array}\right).
\end{equation}
The explicit form for $\epsilon$ is
\begin{eqnarray}
&&\epsilon=\left(\begin{array}{ccc}\epsilon_{11} & 0 & \epsilon_{13} \\
 0 & \epsilon_{2} &0 \\ \epsilon_{13} & 0 & \epsilon_{33}\\\end{array}\right),\nonumber\\
&&\epsilon_{11}=\epsilon_1\cos^2\phi+\epsilon_3\sin^2\phi,\nonumber\\
&&\epsilon_{33}=\epsilon_1\sin^2\phi+\epsilon_3\cos^2\phi,\nonumber\\
&&\epsilon_{13}=\left(\epsilon_3-\epsilon_1\right)\sin\phi\cos\phi.
\end{eqnarray}

\begin{figure}[htbp]
\centering\includegraphics[width=7cm]{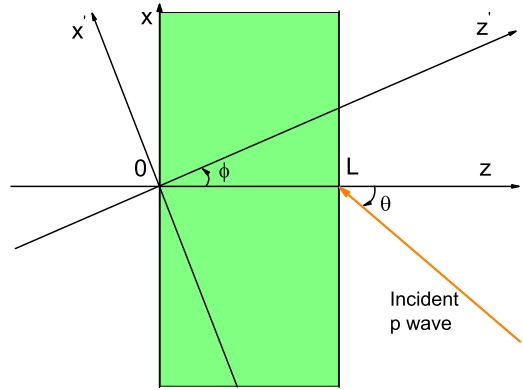}
\caption{Schematic view of an anisotropic medium slab of thickness $L$
with the optical axis $z^\prime$ tilted from the $z$ axis by an angle $\phi$.
A $p$ wave is incident from the right-hand side with an incident angle $\theta$.}
\end{figure}

The anisotropic medium is stratified along the $z$ axis and all tensor components $\epsilon_1$, $\epsilon_2$ and $\epsilon_3$ are arbitrary functions of $z$ in general.
Plane monochromatic EM waves of frequency $\omega$ and vacuum wave number $k_0$ ($=\omega/c$) are
assumed to propagate in the $xz$ plane. Then the wave equations for $s$ and $p$ waves are completely decoupled.
For $p$ waves, the $y$ component of the magnetic field and the $x$ component of the electric field
satisfy the differential equation
\begin{eqnarray}
\frac{d}{dz}\left(\begin{array}{c}H_y\\ E_x\\\end{array}\right)=\left(\begin{array}{cc}-iq\frac{\epsilon_{13}}{\epsilon_{33}} & ik_0\frac{\epsilon_1\epsilon_3}{\epsilon_{33}}\\
ik_0\left(1-\frac{q^2}{{k_0}^2}\frac{1}{\epsilon_{33}}\right) & -iq\frac{\epsilon_{13}}{\epsilon_{33}}\\\end{array}\right)\left(\begin{array}{c}H_y\\ E_x\\\end{array}\right),
\label{eq:pw}
\end{eqnarray}
where $q$ is the $x$ component of the wave vector.
For $s$ waves, the $y$ component of the electric field satisfies
\begin{eqnarray}
\frac{d^2 E_y}{dz^2}+\left({k_0}^2\epsilon_2-q^2\right)E_y=0.
\end{eqnarray}
We can transform Eq.~(\ref{eq:pw}) into a second-order differential equation for $H_y$ by eliminating $E_x$,
though the present form is more convenient for applying the invariant imbedding method.
When $\phi$ is equal to zero, the equation for $H_y$ takes a simple form given by
\begin{eqnarray}
\frac{d^2H_y}{dz^2}-\frac{1}{\epsilon_1}\frac{d\epsilon_1}{dz}\frac{d H_y}{dz}+\left({k_0}^2\epsilon_1
-q^2\frac{\epsilon_1}{\epsilon_3}\right)H_y=0.
\label{eq:hy}
\end{eqnarray}

We assume that an inhomogeneous anisotropic medium
of thickness $L$ lies in $0\le z\le L$ and the waves are incident from a uniform dielectric region ($z>L$)
and transmitted to another uniform dielectric region ($z<0$). The incident and transmitted regions are filled with ordinary isotropic dielectric
media, where $\epsilon$ ($=\epsilon_i$) is a scalar quantity.
In this paper, we are interested only in the resonant absorption phenomenon
associated with the mode conversion of transverse EM waves into longitudinal plasma oscillations. In nonmagnetic media, mode conversion can occur only for $p$ waves. Therefore we need to consider only Eq.~(\ref{eq:pw})
and the configurations for $\epsilon_1$ and $\epsilon_3$.
The tensor component $\epsilon_2$ is completely irrelevant.

\begin{figure}[htbp]
\centering\includegraphics[width=7cm]{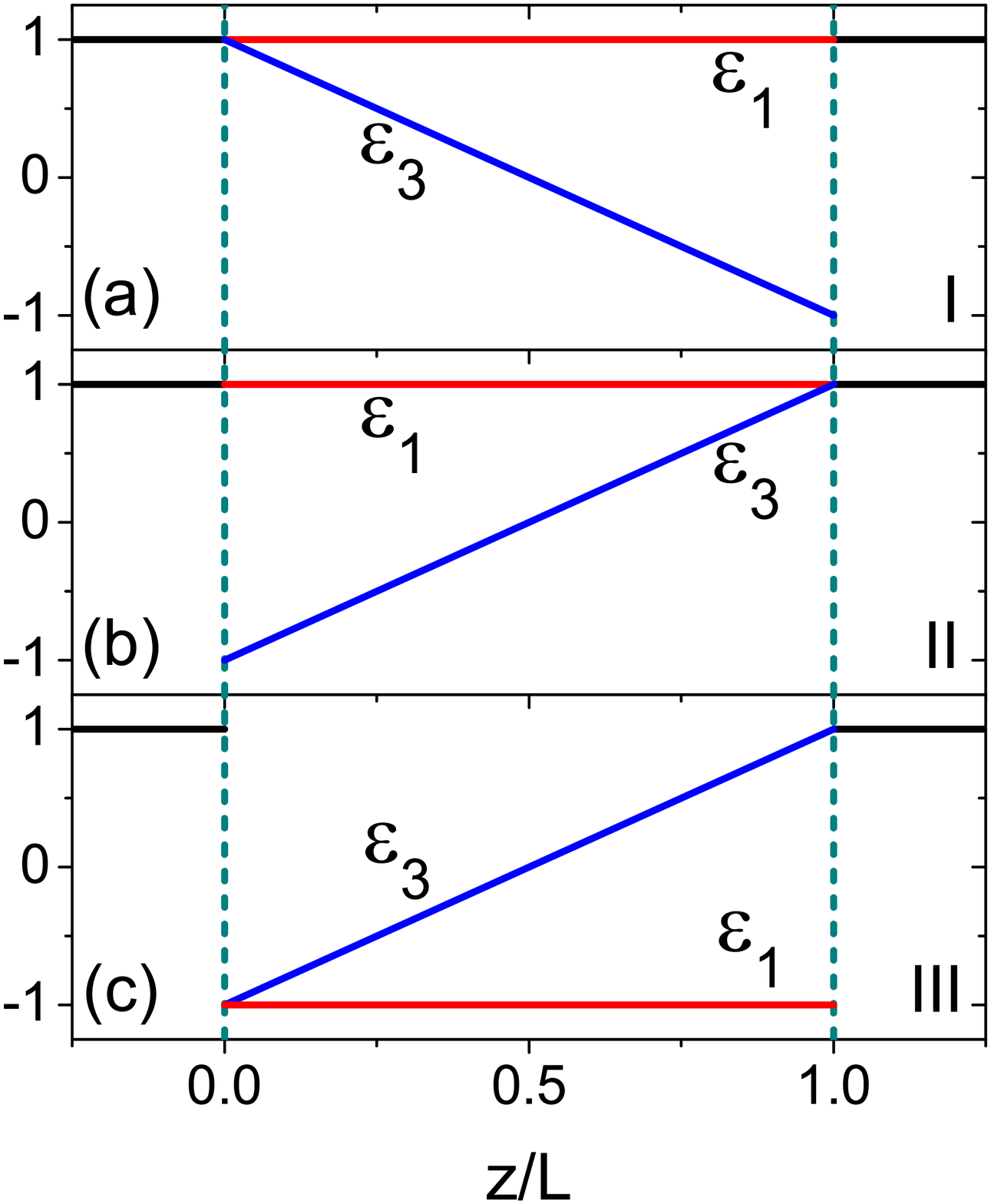}
\caption{Spatial configurations of $\epsilon_1$ and $\epsilon_3$ inside the slab of an anisotropic medium of thickness $L$ considered in this paper.
A $p$ wave is assumed to be incident from the region where $z>L$. In numerical calculations, we introduce extremely small imaginary parts of $\epsilon_1$ and $\epsilon_3$.
\label{fig2}}
\end{figure}

In this work, we consider three different configurations for $\epsilon_1$ and $\epsilon_3$, which are depicted in Fig.~\ref{fig2}.
Mode conversion is found to occur at the positions where $\epsilon_{33}$ ($=\epsilon_1\sin^2\phi+\epsilon_3\cos^2\phi$) is zero
and a singularity occurs in Eq.~(\ref{eq:pw}).
For the purpose of numerical calculations, we introduce extremely small imaginary parts of $\epsilon_1$ and $\epsilon_3$
to avoid the singularity.
Even in the limit where these imaginary parts go to zero,
the absorption of the wave energy is finite due to mode conversion
and converges to a constant. We have verified that the choice of $10^{-8}$ for the imaginary parts yields
sufficiently convergent results for the absorptance. As we mentioned already, mode conversion does not occur for $s$ waves.
In the special case where the tilt angle $\phi$ is zero, $\epsilon_{33}$ is equal to $\epsilon_3$, and therefore
only $\epsilon_3$ plays a role in causing resonant absorption and $\epsilon_1$ is irrelevant. The vanishing of $\epsilon_1$ alone
does not cause resonant absorption.

\section{Invariant imbedding method}
\label{sec3}

We consider a $p$ wave of unit magnitude incident obliquely at an incident angle $\theta$
on the anisotropic medium. In the invariant imbedding method, we are mainly interested in the reflection
and transmission coefficients, $r=r(L)$ and $t=t(L)$, defined by
\begin{eqnarray}
&&H_y(z)=\left\{ \begin{array}{ll}
e^{ip(L-z)}+re^{ip(z-L)},    &z>L \\
te^{-ip z},    &z<0  \end{array} \right.,
\end{eqnarray}
where $p$ ($=\sqrt{\epsilon_i}k_0\cos\theta$) is the negative $z$ component of the wave vector
in the incident and transmitted regions.
Using the invariant imbedding method [27--31], we derive exact differential
equations satisfied by $r$ and $t$:
%\begin{widetext}
\begin{eqnarray}
&&\frac{1}{ip}{{dr}\over{dl}}=2\frac{\epsilon_1\epsilon_3}{\epsilon_{33}\epsilon_i} r+\frac{1}{2}\left(\sec^2\theta-\frac{\epsilon_1\epsilon_3}{\epsilon_{33}\epsilon_i}-\frac{\epsilon_i}{\epsilon_{33}}\tan^2\theta\right)\left(1+r\right)^2,
\nonumber\\
&&\frac{1}{ip}{{dt}\over{dl}}=\left(\frac{\epsilon_1\epsilon_3}{\epsilon_{33}\epsilon_i}+\frac{\epsilon_{13}}{\epsilon_{33}}\tan\theta\right) t+\frac{1}{2}\left(\sec^2\theta-\frac{\epsilon_1\epsilon_3}{\epsilon_{33}\epsilon_i}-\frac{\epsilon_i}{\epsilon_{33}}\tan^2\theta\right)\left(1+r\right)t,
\label{eq:pp}
\end{eqnarray}
%\end{widetext}
starting from Eq.~(\ref{eq:pw}).
These equations are integrated numerically from $l=0$ to $l=L$ using the initial
conditions, $r(0)= 0$ and $t(0)= 1$.

The invariant imbedding method can also be used in
calculating the field amplitude $H_y(z;L)$ inside the inhomogeneous
medium, which is considered as a function of both $z$ and $L$ and satisfies
%\begin{widetext}
\begin{eqnarray}
\frac{1}{ip}{{\partial H_y}\over{\partial l}}=\left(\frac{\epsilon_1\epsilon_3}{\epsilon_{33}\epsilon_i}+\frac{\epsilon_{13}}{\epsilon_{33}}\tan\theta\right) H_y+\frac{1}{2}\left(\sec^2\theta-\frac{\epsilon_1\epsilon_3}{\epsilon_{33}\epsilon_i}-\frac{\epsilon_i}{\epsilon_{33}}\tan^2\theta\right)\left(1+r\right)H_y.
\label{eq:pf}
\end{eqnarray}
%\end{widetext}
For a given $z$ ($0 < z < L$), the field amplitude
is obtained by integrating this equation from $l = z$ to
$l = L$ using the initial condition $H_y(z; z) = 1 + r(z)$.

We notice that when $\phi$ is nonzero (hence $\epsilon_{13}\ne 0$), the equations for the field amplitude and the transmission coefficient are manifestly asymmetric
under the sign change of $\theta$ due to the $\epsilon_{13}\tan\theta$ term, while the equation for the reflection coefficient is always symmetric.
In the next section, we will show that this can lead to asymmetric transmittance and absorptance under the sign change of $\theta$.
Furthermore, we point out that all quantities are symmetric under the simultaneous sign changes of $\theta$ and $\phi$.

The symmetry of the reflection coefficient under the sign change of $\theta$ is a direct consequence of the reciprocity principle, which can be applied to systems with time-reversal symmetry \cite{32}. We note that the reflection coefficients for $\theta$ and $-\theta$ correspond to two time-reversed processes. In contrast to the reflection coefficient, the transmission coefficients for $\theta$ and $-\theta$
do not describe time-reversed processes and are generally unequal. On the other hand, the two transmission coefficients for waves
incident from the right-hand side of the slab and from the left-hand side of the slab in precisely opposite directions correspond to two time-reversed processes and should be the same. Since the configuration II is obtained by flipping the left and right sides of the configuration I, this will lead to the identity of the
transmittances {\it at the same $\theta$} for the configurations I and II.

\section{Numerical results}

\subsection{$\phi=0$ case}

\begin{figure}[htbp]
\centering\includegraphics[width=8cm]{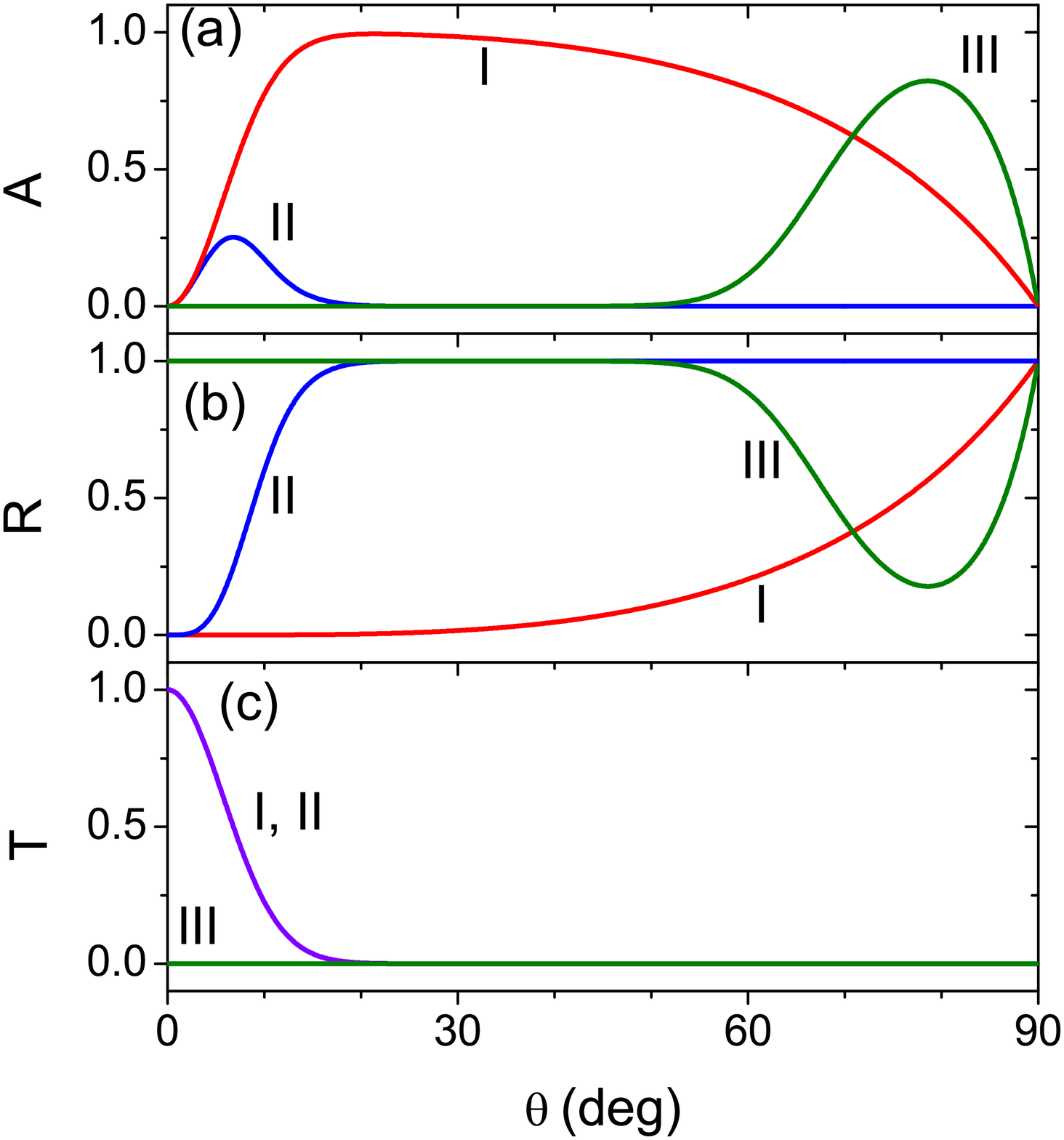}
\caption{(a) Absorptance, (b) reflectance and (c) transmittance of $p$ waves incident from the region where $z>L$ versus incident angle for the configurations I, II and III shown in Fig.~2, when $\phi=0$, $\epsilon_i=1$, ${\rm Im}~\epsilon_3=10^{-8}$ and $L=5\lambda$.}
\end{figure}

We first consider the case where the tilt angle $\phi$ is zero.
Then $H_y$ satisfies Eq.~(\ref{eq:hy}), which can be transformed further by introducing
a new field $\psi$ defined by
\begin{eqnarray}
H_y=\sqrt{\epsilon_1}\psi
\end{eqnarray}
into
\begin{eqnarray}
\psi^{\prime\prime}+\left[k_0^2\epsilon_1-q^2\frac{\epsilon_1}{\epsilon_3}+\frac{1}{2}\frac{{\epsilon_1}^{\prime\prime}}
{\epsilon_1}-\frac{3}{4}\frac{\left({\epsilon_1}^\prime\right)^2}{{\epsilon_1}^2}\right]\psi=0,
\end{eqnarray}
where a prime denotes a differentiation with respect to $z$.
This equation can be written as
\begin{eqnarray}
\psi^{\prime\prime}+k^2\left[1-\eta(z)\right]\psi=0,
\label{eq:f1}
\end{eqnarray}
where $k=\sqrt{\epsilon_i}k_0$ and
$\eta$ is given by
\begin{eqnarray}
\eta = 1-\frac{\epsilon_1}{\epsilon_i}+\frac{\epsilon_1}{\epsilon_3}\sin^2\theta
 -\frac{1}{2}\frac{1}{k^2}\frac{{\epsilon_1}^{\prime\prime}}
{\epsilon_1}+\frac{3}{4}\frac{1}{k^2}\frac{\left({\epsilon_1}^\prime\right)^2}{{\epsilon_1}^2}.
\label{eq:f2}
\end{eqnarray}
Equation~(\ref{eq:f1}) has the same form as the Schr\"odinger equation and the function $\eta$ plays the role of $V(z)/E$,
where $V(z)$ is the potential and $E$ is the energy of an incident quantum particle.
For given configurations of $\epsilon_1$ and $\epsilon_3$, it is possible to get some insights on the wave propagation by looking at the form of the function $\eta(z)$.

For the configurations considered in this paper and when $\phi$ is zero, mode conversion occurs at $z=0.5L$, where
${\rm Re}~\epsilon_3$ vanishes. In order to have strong resonant absorption via mode conversion,
it is necessary for a large fraction of wave energy to arrive at $z=0.5L$
without substantial reflection by the medium in $0.5<z/L<1$. Strong reflection can result from the presence of an evanescent region, where $\eta>1$
and the wave number is complex.
Therefore, if there exists an evanescent region of substantial size in $0.5<z/L<1$, we expect mode conversion to be strongly suppressed.

In Fig.~3, we plot the reflectance $R$ ($=\vert r\vert^2$), the transmittance $T$ ($=\vert t\vert^2$) and the absorptance $A$ ($=1-R-T$), which can be equivalently called the mode conversion coefficient,
for $p$ waves incident on the slabs described by the configurations I, II and III shown in Fig.~2 versus incident angle, when $\phi=0$, $\epsilon_i=1$, ${\rm Im}~\epsilon_3=10^{-8}$ and $L=5\lambda$. We find that the absorptance is fairly large in a wide range of the incident angle
in Case I, while it is substantial only in the small $\theta$ region in Case II and in the large $\theta$ region in Case III. The maximum values of
$A$ are 0.994 at $\theta=21.4^\circ$ in Case I, 0.252 at $\theta=6.8^\circ$ in Case II and 0.822 at $\theta=78.6^\circ$ in Case III. We note that
the absorptance in Case II is substantially smaller than those in the other cases.

In Case I, the transverse and longitudinal tensor components, $\epsilon_1$ and $\epsilon_3$, have different signs in the region $0.5<z/L<1$, where the slab can be considered as an inhomogeneous hyperbolic medium. More specifically, since $\epsilon_3$ is negative, it can be regarded as a type I hyperbolic medium, if $\epsilon_2=\epsilon_1$.
The function $\eta$ is given by
\begin{eqnarray}
\eta_{\rm I} = -\frac{\sin^2\theta}{2\left(z/L\right)-1}.
\end{eqnarray}
Since $\eta_{\rm I}$ is negative in $0.5<z/L<1$, no evanescent region occurs there for any incident angle, which results in a large absorption
over a wide range of $\theta$, except for $\theta$ very close to $0^\circ$ or $90^\circ$.
The reflectance increases monotonically from 0 to 1 as $\theta$ increases,
which leads to a suppression of $A$ for large $\theta$. This suppression results from two factors.
First, the mode conversion efficiency decreases as $\theta$ increases to $90^\circ$, which makes a larger fraction
of the wave energy reach beyond the resonance region at $z=0.5L$. Second, an evanescent region occurs in $0.5\cos^2\theta<
z/L<0.5$. The width of this region increases, and therefore the reflectance increases and the transmittance decreases, as $\theta$ increases. When $\theta$ is close to zero,
the effective potential $\eta$ is very small, since it is proportional to $\sin^2\theta$. This makes the slab almost transparent, with $T\approx 1$ and $A\approx 0$.

In Case II, $\epsilon_1$ and $\epsilon_3$ have the same signs in $0.5<z/L<1$, where the slab is an elliptic medium.
The function $\eta$ is given by
\begin{eqnarray}
\eta_{\rm II} = -\eta_{\rm I}=\frac{\sin^2\theta}{2\left(z/L\right)-1}.
\end{eqnarray}
An evanescent region appears in $0.5<z/L<0.5(1+\sin^2\theta)$. Since the width of this region increases rapidly as $\theta$ increases,
total reflection and zero absorption occur for reasonably large values of $\theta$. When $\theta$ is close to zero, $\eta$ is very small, which results in a large transmission and a small absorption, as in Case I. As we have already explained in Sec.~3,
the transmittances in Case I and II are identically the same because of the reciprocity principle.

In Case III, $\epsilon_1$ and $\epsilon_3$ have different signs in $0.5<z/L<1$, where the slab is a hyperbolic medium. Since $\epsilon_1$ is negative, it can be regarded as a type II hyperbolic medium, if $\epsilon_2=\epsilon_1$.
The function $\eta$ is given by
\begin{eqnarray}
\eta_{\rm III} = 2-\frac{\sin^2\theta}{2\left(z/L\right)-1}.
\end{eqnarray}
In this case, the wave is evanescent everywhere except in $0.5<z/L<0.5(1+\sin^2\theta)$. In order to have nonzero absorption, the width of
this region has to be sufficiently large. For $\theta$ smaller than about $50^\circ$, the slab behaves as a perfect reflector and shows
zero absorption. Since the wave is evanescent in the wide region $0<a/L<0.5$ for any incident angle,
the transmittance is always zero.

Summarizing the discussions above, we conclude that mode conversion is
much more enhanced in the case where the medium from the incident surface to the resonance region is hyperbolic than that where
it is elliptic. A type I hyperbolic medium is more efficient in this regard than a type II hyperbolic medium.

\begin{figure}[htbp]
\centering\includegraphics[width=8cm]{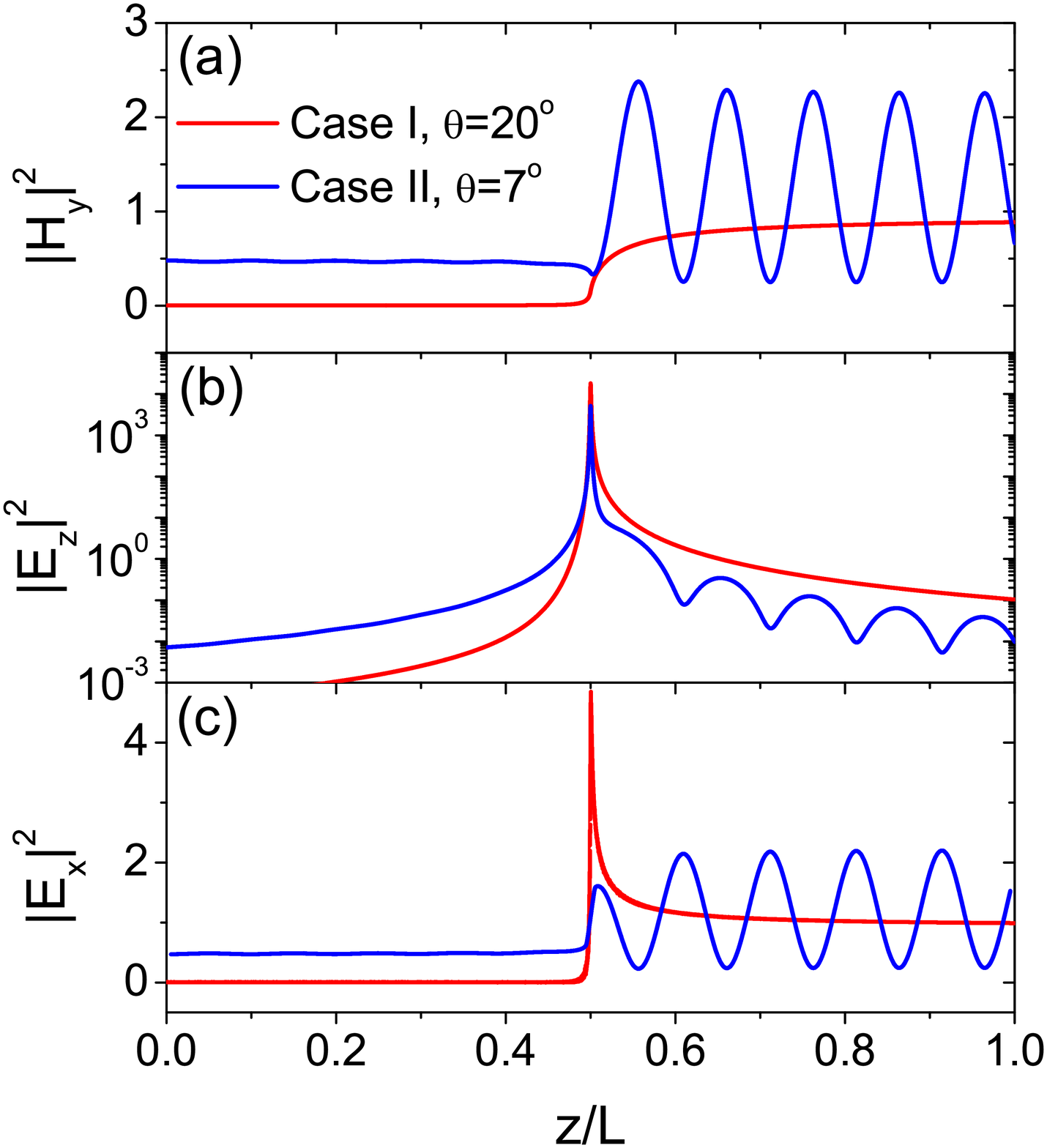}
\caption{Spatial distributions of (a) the magnetic field, (b) the $z$ component of the electric field and (c) the $x$ component of the electric field inside the inhomogeneous anisotropic medium when a $p$ wave is incident from the region where $z>L$ on the slab I of Fig.~2 at $\theta=20^\circ$ and on the slab II of Fig.~2 at $\theta=7^\circ$, when $\phi=0$, $\epsilon_i=1$, ${\rm Im}~\epsilon_3=10^{-3}$ and $L=5\lambda$.}
\end{figure}

In Fig.~4, we show two examples of the spatial distributions of the electric and magnetic fields inside the inhomogeneous slab when a $p$ wave is incident from the region where $z>L$. When the tilt angle $\phi$ is zero, the electric field components are obtained using
\begin{equation}
E_z=-\frac{q}{k_0}\frac{H_y}{\epsilon_3},~~~E_x=-\frac{i}{k_0\epsilon_1}\frac{dH_y}{dz}.
\end{equation}
We consider the cases where a wave is incident on the slab I of Fig.~2 at $\theta=20^\circ$ and on the slab II of Fig.~2 at $\theta=7^\circ$, when $\phi=0$, $\epsilon_i=1$, ${\rm Im}~\epsilon_3=10^{-3}$ and $L=5\lambda$. In the former case, $R$ ($=0.0034$) is very small and $A$ ($=0.993$) is large,
while, in the latter, $R$ is 0.268 and $A$ is 0.252.
We observe clearly that mode conversion of transverse waves into longitudinal plasma oscillations occurs at $z=0.5L$ where
${\rm Re}~\epsilon_z=0$, as can be seen from the strong enhancement of $E_z$ near $z=0.5L$ in Fig.~4(b).
When the reflectance is not very small in Case II, the incident waves are primarily reflected from the boundary of the evanescent region located at $z=0.5(1+\sin^2\theta)L$. The wave incident from the right and propagating to the left and that reflected from the evanescent region interfere and the field
distribution shows oscillatory behavior in the region $0.5(1+\sin^2\theta)L<z<L$, as can be clearly seen in the blue curves. If the reflectance is negligible, no oscillatory feature is observed as in the red curves.

\subsection{$\phi\ne 0$ case}

\begin{figure}[htbp]
\centering\includegraphics[width=9cm]{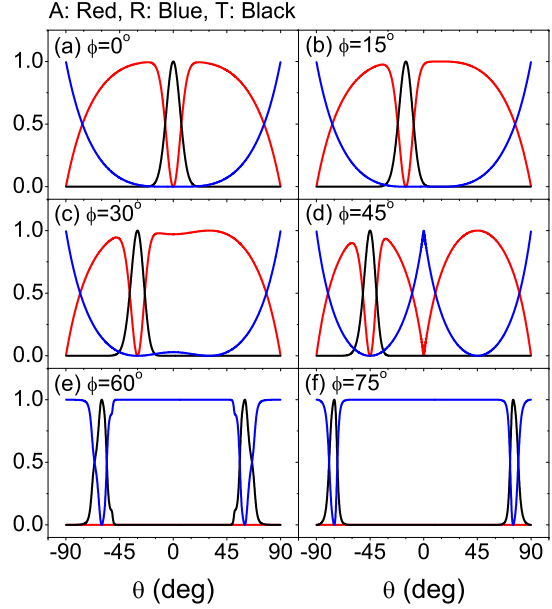}
\caption{Absorptance (red), reflectance (blue) and transmittance (black) of $p$ waves incident from the region where $z>L$ versus incident angle for the configuration I shown in Fig.~2, when $\epsilon_i=1$, ${\rm Im}~\epsilon_3=10^{-8}$, $L=5\lambda$ and (a) $\phi=0^\circ$, (b) $15^\circ$, (c) $30^\circ$,
(d) $45^\circ$, (e) $60^\circ$ and (f) $75^\circ$.}
\end{figure}

Next, we consider the more general case where the tilt angle $\phi$ is nonzero. In Fig.~5, we show the absorptance, the reflectance and the transmittance versus incident angle for the configuration I in Fig.~2, when $\epsilon_i=1$, ${\rm Im}~\epsilon_3=10^{-8}$, $L=5\lambda$ and $\phi=0^\circ$, $15^\circ$, $30^\circ$,
$45^\circ$, $60^\circ$ and $75^\circ$. As we have already mentioned in Sec.~\ref{sec3}, all quantities are symmetric under the
simultaneous sign changes of $\theta$ and $\phi$. Therefore, we need to consider only the cases with $\phi>0$. We find that the absorptance is zero for all incident angles, if $\phi>45^\circ$. This can be understood easily from the form of $\epsilon_{33}$ given by
\begin{eqnarray}
\epsilon_{33}=1-2\frac{z}{L}\cos^2\phi.
\end{eqnarray}
The resonance point $z_{\rm R}$ at which mode conversion occurs is determined by $\epsilon_{33}=0$, which gives
\begin{eqnarray}
\frac{z_{\rm R}}{L}=\frac{1}{2\cos^2\phi}.
\end{eqnarray}
If $\phi>45^\circ$, this condition cannot be satisfied at any point within the medium.

Another interesting feature is that both $A$ and $T$ are asymmetric under the sign change of $\theta$ when $0<\phi\le 45^\circ$, whereas
$T$ is symmetric and $A$ is zero when $\phi> 45^\circ$. Since $A$ is zero for all $\theta$ when $\phi> 45^\circ$,
the symmetry of $T$ has to follow from the law of
energy conservation $R+T+A=R+T=1$, considering that $R$ is always symmetric.
When $\phi\le 45^\circ$, the transmittance curve is sharply peaked at $\theta= -\phi$, with the peak value equal to 1.
This corresponds to the situation where the wave is incident precisely along the optical axis $z^\prime$.
When $\phi> 45^\circ$, the transmittance curve has two peaks at $\theta= \pm\phi$.
This can be understood from the behavior of the coefficient of the $(1+r)^2$ and $(1+r)t$ terms in Eq.~(\ref{eq:pp}):
\begin{eqnarray}
\sec^2\theta-\frac{\epsilon_1\epsilon_3}{\epsilon_{33}\epsilon_i}-\frac{\epsilon_i}{\epsilon_{33}}\tan^2\theta=
\frac{2(z/L)\left(\cos^2\theta-\cos^2\phi\right)}{\epsilon_{33}\cos^2\theta},
\end{eqnarray}
which vanishes at $\theta=\pm\phi$.
When $\theta=-\phi$, the inhomogeneous slab is effectively a transparent medium with $T=1$ and $R=A=0$ for all values of $\phi$.
In contrast, when $\theta=\phi$, the slab is transparent with $T=1$ and $R=A=0$ for $\phi> 45^\circ$, whereas it is a perfect absorber
with $A=1$ and $R=T=0$ for $0<\phi\le 45^\circ$.

As $\phi$ increases from zero to $45^\circ$, the reflectance near $\theta=0$ increases monotonically. For $\phi> 45^\circ$, the slab is a perfect reflector at all incident angles except for those in a narrow region around $\theta=\pm\phi$. When $\phi$ is equal to $90^\circ$, the permittivity configuration is the same as that for the $\phi=0$ case with $\epsilon_3=1$ and
$\epsilon_1=1-2(z/L)$. The effective potential $\eta$ for this configuration is given by
\begin{eqnarray}
\eta=1+\left(2\frac{z}{L}-1\right)\cos^2\theta+\frac{3}{\zeta^2\left[2(z/L)-1\right]^2},
\end{eqnarray}
where $\zeta=kL=10\pi$.
Since this function is always larger than 1 in $0.5<z/L<1$, the wave is evanescent in that wide region, which causes the slab to
behave as a total reflector at all incident angles.

\begin{figure}[htbp]
\centering\includegraphics[width=9cm]{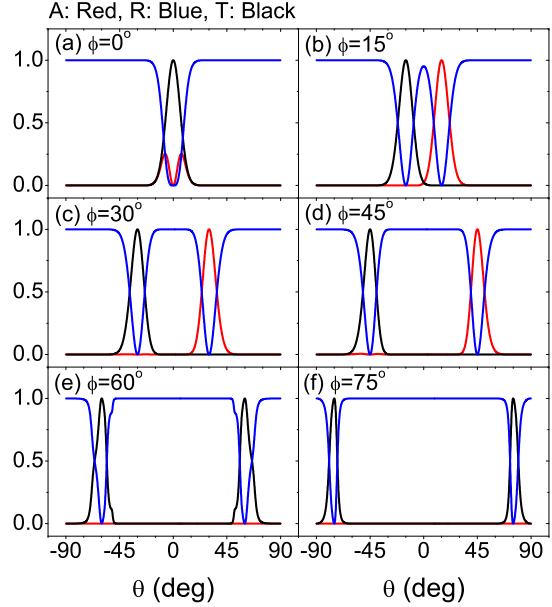}
\caption{Absorptance (red), reflectance (blue) and transmittance (black) of $p$ waves incident from the region where $z>L$ versus incident angle for the configuration II in Fig.~2, when $\epsilon_i=1$, ${\rm Im}~\epsilon_3=10^{-8}$, $L=5\lambda$ and (a) $\phi=0^\circ$, (b) $15^\circ$, (c) $30^\circ$,
(d) $45^\circ$, (e) $60^\circ$ and (f) $75^\circ$.}
\end{figure}

In Fig.~6, we show the absorptance, the reflectance and the transmittance
versus incident angle for the configuration II shown in Fig.~2, for the same parameter values as in Fig.~5.
Similarly to Fig.~5, we find that the absorptance is zero for all incident angles, if $\phi>45^\circ$. The function $\epsilon_{33}$ in the present case is given by
\begin{eqnarray}
\epsilon_{33}=1+2\left(\frac{z}{L}-1\right)\cos^2\phi.
\end{eqnarray}
The resonance point $z_{\rm R}$ satisfies
\begin{eqnarray}
\frac{z_{\rm R}}{L}=1-\frac{1}{2\cos^2\phi}.
\end{eqnarray}
If $\phi$ is greater than $45^\circ$, this condition cannot be satisfied at any point within the medium, and therefore $A$ is always zero..
The curves for the transmittance, which show one peak at $\theta=-\phi$ if $\phi\le45^\circ$ and two peaks at
$\theta=\pm\phi$ if $\phi>45^\circ$, are identical to those in Fig.~5 for all values of $\phi$, which is a consequence of the reciprocity principle.
Except when the incident angle is near $\pm\phi$, the slab behaves as a perfect reflector with $R= 1$. For very small values of $\phi$, the absorptance curve shows two small peaks near $\theta=0$, but for $\phi$ larger than about $5^\circ$, it has one large peak at $\theta=\phi$
with the peak value equal to 1.

\begin{figure}[htbp]
\centering\includegraphics[width=9cm]{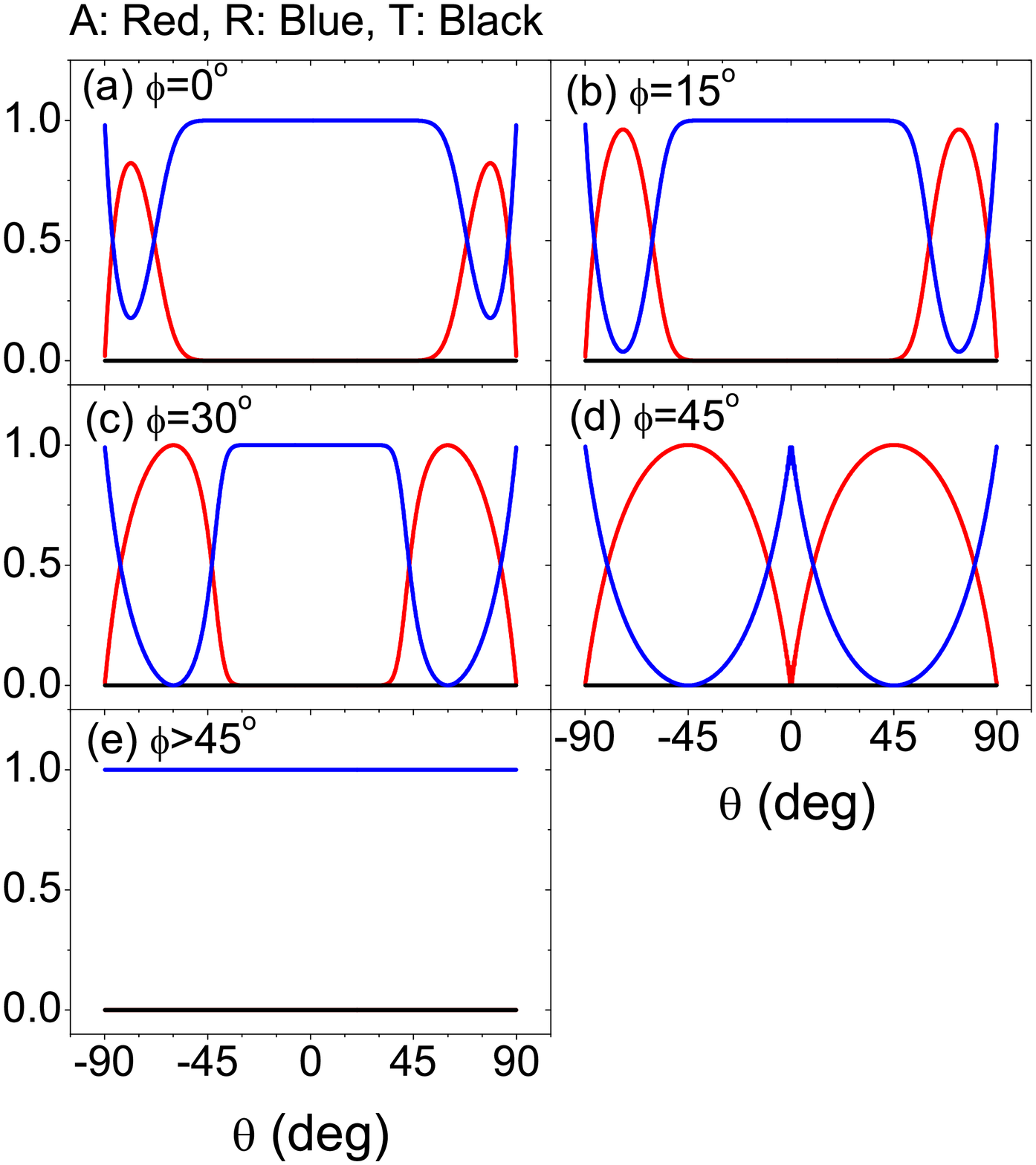}
\caption{Absorptance (red), reflectance (blue) and transmittance (black) of $p$ waves incident from the region where $z>L$ versus incident angle for the configuration III in Fig.~2, when $\epsilon_i=1$, ${\rm Im}~\epsilon_3=10^{-8}$, $L=5\lambda$ and (a) $\phi=0^\circ$, (b) $15^\circ$, (c) $30^\circ$,
(d) $45^\circ$ and (e) $\phi>45^\circ$. In (e), both $T$ and $A$ are zero.}
\end{figure}

In Fig.~7, we show the absorptance, the reflectance and the transmittance
versus incident angle for the configuration III in Fig.~2, when $\epsilon_i=1$, ${\rm Im}~\epsilon_3=10^{-8}$, $L=5\lambda$ and $\phi=0^\circ$, $15^\circ$, $30^\circ$,
$45^\circ$ and $\theta>45^\circ$.
Once again, we find that the absorptance is zero for all incident angles, if $\phi>45^\circ$. $\epsilon_{33}$ in the present case is given by
\begin{eqnarray}
\epsilon_{33}=2\frac{z}{L}\cos^2\phi-1.
\end{eqnarray}
The equation for the resonance point $z_{\rm R}$ is the same as Eq.~(19) and cannot be satisfied at any point within the medium if $\phi>45^\circ$,
which leads to the vanishing of $A$.
In the present case, the evanescent region is so wide that there is no transmission at any incident angle for all values of $\phi$.
Since $R$ is always symmetric under the sign change of $\theta$, the absorptance is also symmetric as shown in Fig.~7.
When $\phi>45^\circ$, both $T$ and $A$ are zero for all $\theta$, and therefore the slab acts as an omnidirectional perfect reflector.
As $\phi$ increases from zero to $45^\circ$, the broad absorption peaks become wider and move to $\theta=\pm 45^\circ$.

\begin{figure}[htbp]
\centering\includegraphics[width=9cm]{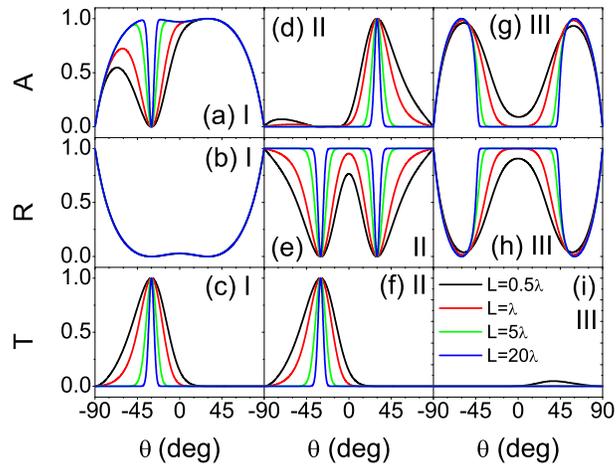}
\caption{Absorptance, reflectance and transmittance versus incident angle for (a-c) Case I, (d-f) Case II and (g-i) Case III, when ${\rm Im}~\epsilon_3=10^{-8}$, $\epsilon_i=1$, $\phi=30^\circ$
 and $L=0.5\lambda$ (black), $\lambda$ (red), $5\lambda$ (green) and $20\lambda$ (blue).}
\end{figure}

\begin{figure}[htbp]
\centering\includegraphics[width=8cm]{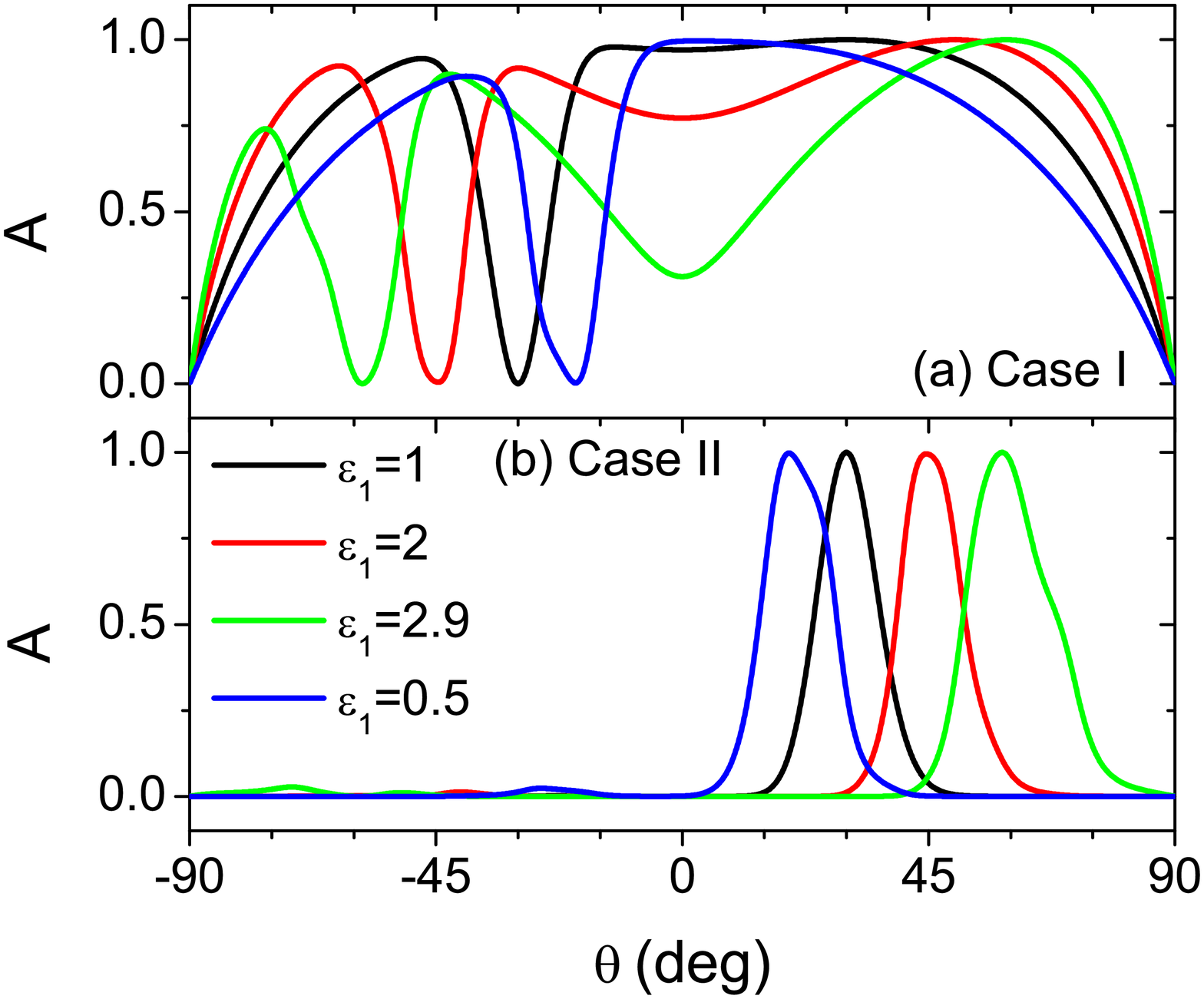}
\caption{Absorptance versus incident angle for the configurations similar to (a) I and
 (b) II in Fig.~2, when ${\rm Im}~\epsilon_3=10^{-8}$, $L=5\lambda$, $\phi=30^\circ$, $\epsilon_i=1$
 and $\epsilon_1=1$ (black), 2 (red), 2.9 (green) and 0.5 (blue).}
\end{figure}

\begin{figure}[htbp]
\centering\includegraphics[width=8cm]{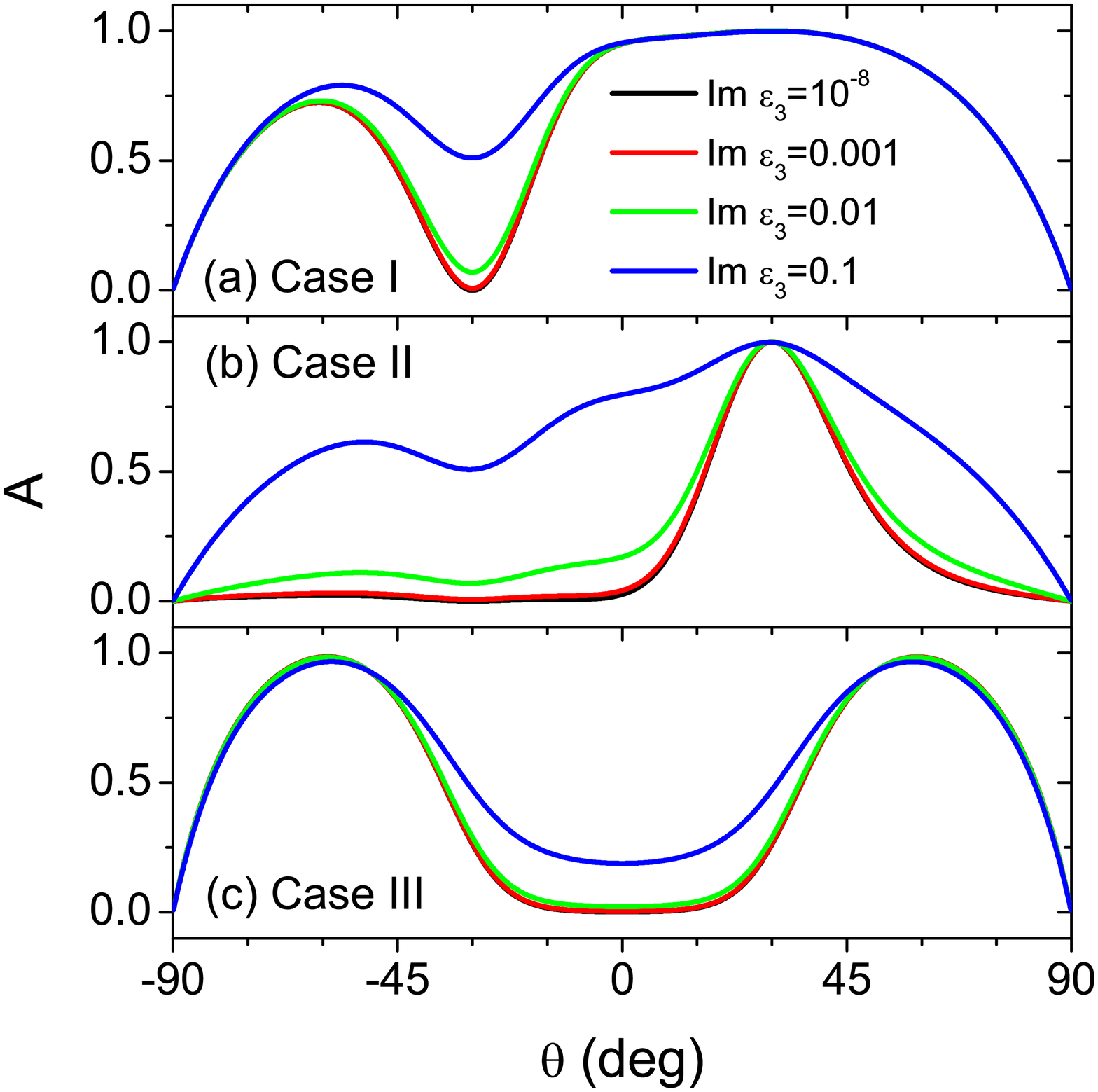}
\caption{Absorptance versus incident angle for (a) Case I,
 (b) Case II and (c) Case III in Fig.~2, when $L=\lambda$, $\phi=30^\circ$,
 $\epsilon_i=1$ and ${\rm Im}~\epsilon_3=10^{-8}$ (black), 0.001 (red), 0.01 (green) and 0.1 (blue).}
\end{figure}

Next, we discuss the effects of the thickness of the slab, the mismatch between $\epsilon_1$ and the background $\epsilon_i$ and the loss parameter ${\rm Im}~\epsilon_3$
on the results presented above. In Fig.~8, we show how the absorptance, reflectance and transmittance curves change as the thickness of the
inhomogeneous layer increases, for the configurations I, II and III, when ${\rm Im}~\epsilon_3=10^{-8}$, $\epsilon_i=1$, $\phi=30^\circ$
 and $L=0.5\lambda$, $\lambda$, $5\lambda$ and $20\lambda$. We notice a general trend that the peaks and dips become sharper as
 $L/\lambda$ increases. Nevertheless, general features of mode conversion phenomena are observable for $L$ as small as $0.5\lambda$.

In Fig.~9, we show how the mismatch between $\epsilon_i$ and $\epsilon_1$ influences the absorptance. The configurations are the same as
I and II in Fig.~2, except that $\epsilon_1$ is generally not equal to 1. The expression for $\epsilon_{33}$ given in Eq.~(4) depends on $\epsilon_1$.
For $\phi=30^\circ$ chosen in Fig.~9, we can show that the region where ${\rm Re}~\epsilon_{33}=0$ can exist only when $0<\epsilon_1<3$ for
both cases. Furthermore, by examining the left-hand side of Eq.~(21) for $\epsilon_1\ne 1$, we can show that the absolute value of the incident angle at which $A=0$
in Case I and $A=1$ in Case II increases (decreases) as $\epsilon_1$ increases (decreases) from 1. Though the detailed shape of the absorptance curve depends on $\epsilon_1$, mode conversion and resonant absorption behaviors are qualitatively similar to the matched case.

In Fig.~10, we consider the influence of the loss parameter, ${\rm Im}~\epsilon_3$, on mode conversion phenomena.
Resonant absorption due to mode conversion is not due to damping, but rather to the conversion of transverse wave modes into longitudinal modes.
Therefore the absorptance in this case converges to a finite value even in the limit where ${\rm Im}~\epsilon_3$ goes to zero.
In realistic cases, however, there is always finite damping, which will enhance the overall absorption. In Fig.~10, we consider a thin slab
of thickness $L=\lambda$ for the configurations I, II and III in Fig.~2 and vary ${\rm Im}~\epsilon_3$. We notice that the results for ${\rm Im}~\epsilon_3=0.001$ are almost the same as those obtained when ${\rm Im}~\epsilon_3\rightarrow 0^+$. The influence of the loss parameter appears to be stronger in Case II than in the other cases. We find that the value of 0.01 is sufficiently good and mode conversion phenomena
are not dominated by damping if ${\rm Im}~\epsilon_3$ is smaller than about 0.01.

\begin{figure}[htbp]
\centering\includegraphics[width=8cm]{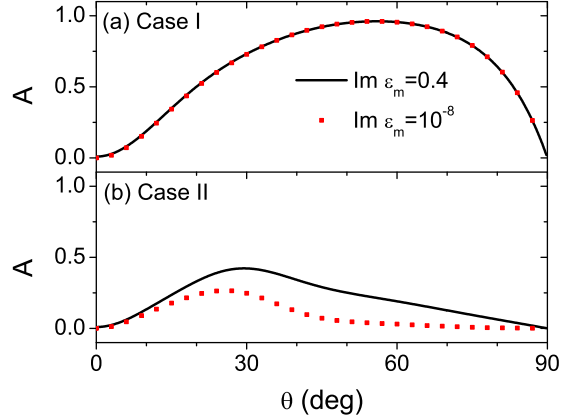}
\caption{Absorptance versus incident angle for the configurations similar to (a) I and
 (b) II in Fig.~2, when $L=\lambda$, $\phi=0^\circ$ and
 $\epsilon_i=1$. The configurations correspond to graded metal-dielectric multilayers made of Silver with
 $\epsilon_m=-15.24+0.4i$ at $\lambda=587.6$ nm and ${\rm MgF_2}$ with $\epsilon_d=1.9$. An effective medium description where
 the metal fraction $f$ varies linearly as $f=0.1+0.2(z/L)$ in (a) and as $f=0.3-0.2(z/L)$ in (b) has been used. The results for ${\rm Im}~\epsilon_m=0.4$ are compared with those obtained in the hypothetical cases with ${\rm Im}~\epsilon_m=10^{-8}$.}
\end{figure}

Finally, we comment on the experimental feasibility of the results presented in this paper. Even though we have chosen rather specific types
of linear configurations for the purpose of simplicity, their detailed shapes are not essential for the occurrence of mode conversion 
and resonant absorption phenomena. What is essential is that there has to a region where the real part of $\epsilon_{33}$ in the tilted case
and $\epsilon_3$ in the non-tilted case vanishes. Transition anisotropic media can be fabricated using
a scheme similar to that for hyperbolic media such as metal-dielectric multilayers or metallic wire arrays \cite{24,25}. In case of metal-dielectric multilayers,
one can use an effective medium description such that
\begin{eqnarray}
\epsilon_3=f\epsilon_m+(1-f)\epsilon_d,~~~\frac{1}{\epsilon_1}=\frac{f}{\epsilon_m}+\frac{1-f}{\epsilon_d},
\end{eqnarray}
where $\epsilon_m$ and $\epsilon_d$ are the dielectric permittivities of the metal and dielectric layers and $f$
is the metal fraction. In Fig.~11, we show the absorptance versus incident angle curve for the configurations similar to I and
 II in Fig.~2, when $L=\lambda$, $\phi=0^\circ$ and
 $\epsilon_i=1$. The configurations correspond to graded metal-dielectric multilayers made of Silver with
 $\epsilon_m=-15.24+0.4i$ at $\lambda=587.6$ nm and ${\rm MgF_2}$ with $\epsilon_d=1.9$. An effective medium description where
 the metal fraction $f$ varies linearly as $f=0.1+0.2(z/L)$ in (a) and as $f=0.3-0.2(z/L)$ in (b) has been used. In the experimental fabrication of these systems, we suggest to fabricate a graded multilayer with 20 periods, where the thickness of each period is 40 nm and the metal fraction 
 is varied from 0.1 to 0.29 with a step
 of 0.01. We notice that the basic features of mode conversion phenomena are clearly observed in Fig.~11. In Fig.~11(a), the obtained result is
 almost the same as in the hypothetical case where the metallic damping is ignored. 

\section{Conclusion}

In this paper, we have studied theoretically the mode conversion and resonant absorption phenomena occurring in an inhomogeneous
anisotropic medium. We have considered the more general asymmetric case, where the optical axes of the medium are tilted with respect to the direction of stratification.
When the tilt angle is zero, we have found that mode conversion occurs if the longitudinal component of the permittivity tensor varies from positive to negative values within the medium, while the transverse component is irrelevant in the mode conversion process.
When the tilt angle is nonzero, we have found that the curves for the transmittance and the absorptance can be asymmetric under the sign change of the incident angle, while the reflectance curve is always symmetric. We have calculated the reflectance,
the transmittance and the absorptance for three configurations of the permittivity tensor, using the invariant imbedding method, and found that resonant absorption is
much more enhanced in the case where the medium from the incident surface to the resonance region is hyperbolic than that where
it is elliptic. We expect that the asymmetric transmission and absorption phenomena discussed in this paper will find
useful applications in photonic devices.

\section*{Funding}
National Research Foundation of Korea Grant funded by the Korean Government (NRF-2015R1A2A2A01003494).


\begin{thebibliography}{99}

\bibitem{1} K. Kim, D.-H. Lee, and H. Lim, ``Resonant absorption and mode conversion in a transition layer between positive-index and negative-index media,'' Opt. Express {\bf 16}(22), 18505--18513 (2008).
\bibitem{2} I. Mozjerin, E. A. Gibson, E. P. Furlani, I. R. Gabitov, and N. M. Litchinitser, ``Electromagnetic enhancement in lossy optical transition metamaterials,'' Opt. Lett. {\bfseries 35}(19), 3240--3242 (2010).
\bibitem{3} E. A. Gibson, M. Pennybacker, A. I. Maimistov, I. R. Gabitov, and N. M. Litchinitser, ``Resonant absorption in transition metamaterials: parametric study,'' J. Opt. {\bf 13}(2), 024013 (2011).
\bibitem{4} J. Sun, X. Liu, J. Zhou, Z. Kudyshev, and N. M. Litchinitser, ``Experimental demonstration of anomalous field enhancement in all-dielectric transition magnetic metamaterials,'' Sci. Rep. {\bfseries 5}, 16154 (2015).
\bibitem{5} S. Kim and K. Kim, ``Resonant absorption and amplification of circularly-polarized waves in inhomogeneous chiral media,''
Opt. Express {\bfseries 24}(2), 1794--1803 (2016).
\bibitem{6} Y. Jin, S. Xiao, N. A. Mortensen, and S. He, ``Arbitrarily thin metamaterial structure for perfect absorption and giant magnification,'' Opt. Express {\bf 19}(12), 11114--11119 (2011).
\bibitem{7} J. Yoon, M. Zhou, M. A. Badsha, T. Y. Kim, Y. C. Jun, and C. K. Hwangbo, ``Broadband epsilon-near-zero perfect absorption in the near-infrared,'' Sci. Rep. {\bf 5}, 12788 (2015).
\bibitem{8} M. Lobet, B. Majerus, L. Henrard, and P. Lambin, ``Perfect electromagnetic absorption using graphene and epsilon-near-zero metamaterials,'' Phys. Rev. B {\bfseries 93}(23), 235424 (2016).
\bibitem{9} S. Feng and K. Halterman, ``Coherent perfect absorption in epsilon-near-zero metamaterials,'' Phys. Rev. B {\bf 86}(16), 165103 (2012).
\bibitem{10} I. S. Nefedov, C. A. Valagiannopoulos, S. M. Hashemi, and E. I. Nefedov, ``Total absorption in asymmetric hyperbolic media,''
Sci. Rep. {\bfseries 3}, 2662 (2013).
\bibitem{11} S. Zhong, Y. Ma, and S. He, ``Perfect absorption in ultrathin anisotropic $\epsilon$-near-zero metamaterials,'' Appl. Phys. Lett. {\bf 105}(2), 023504 (2014).
\bibitem{12} K. Halterman and J. M. Elson, ``Near-perfect absorption in epsilon-near-zero structures with hyperbolic dispersion,'' Opt. Express {\bf 22}(6), 7337--7348 (2014).
\bibitem{13} J. Linder and K. Halterman, ``Graphene-based extremely wideangle tunable metamaterial absorber,'' Sci. Rep. {\bfseries 6}, 31225 (2016).
\bibitem{14} K. V. Sreekanth, M. ElKabbash, Y. Alapan, A. R. Rashed, U. A. Gurkan, and G. Strangi, ``A multiband perfect absorber based
on hyperbolic metamaterials,'' Sci. Rep. {\bfseries 6}, 26272 (2016).
\bibitem{15} D. G. Swanson,
{\it Theory of Mode Conversion and Tunneling in Inhomogeneous
Plasmas} (Wiley, 1998).
\bibitem{16} D. E. Hinkel-Lipsker, B. D. Fried, and G. J.
Morales, ``Analytic expressions for mode conversion in a plasma with
a linear density profile,'' Phys. Fluids B {\bf 4}(3), 559--575 (1992).
\bibitem{17} E. Mj{\o}lhus, ``On linear conversion in a magnetized plasma,'' Radio Sci. {\bf 25}(6), 1321--1339 (1990).
\bibitem{18}
K. Kim and D.-H. Lee, ``Invariant imbedding theory of mode
conversion in inhomogeneous plasmas. I. Exact calculation of the
mode conversion coefficient in cold, unmagnetized plasmas,'' Phys.
Plasmas {\bf 12}(6), 062101 (2005).
\bibitem{19} K. Kim and D.-H. Lee, ``Invariant imbedding theory of mode conversion in inhomogeneous
plasmas. II. Mode conversion in cold, magnetized plasmas with
perpendicular inhomogeneity,'' Phys. Plasmas {\bf 13}(4), 042103
(2006).
\bibitem{20} D. J. Yu, K. Kim, and D.-H. Lee, ``Resonant enhancement of mode conversion in unmagnetized plasmas due to a periodic density modulation superimposed on a linear electron density profile,'' Phys. Plasmas {\bf 17}(10), 102110 (2010).
\bibitem{21} D. J. Yu, K. Kim, and D.-H. Lee, ``Temperature dependence of mode conversion in warm, unmagnetized plasmas with a linear density profile,'' Phys. Plasmas {\bf 20}(6), 062109 (2013).
\bibitem{22} D. J. Yu and K. Kim, ``Effects of a random spatial variation of the plasma density on the mode conversion in cold, unmagnetized, and stratified plasmas,'' Phys. Plasmas {\bf 20}(12), 122104 (2013).
\bibitem{23} D. J. Yu and K. Kim, ``Broadband wide-angle absorption enhancement due to mode conversion in cold
unmagnetized plasmas with periodic density variations,'' Phys. Plasmas {\bf 23}(3), 032112 (2016).
\bibitem{24} A. Poddubny, I. Iorsh, P. Belov, and Y. Kivshar, ``Hyperbolic metamaterials,'' Nat. Photon. {\bf 7}(12), 948--957 (2013).
\bibitem{25} L. Ferrari, C. Wu, D. Lepage, X. Zhang, and Z. Liu, ``Hyperbolic metamaterials and their applications,''
Prog. Quantum Electron. {\bfseries 40}, 1--40 (2015).
\bibitem{26} G. Pawlik, K. Tarnowski, W. Walasik, A. C. Mitus, and I. C. Khoo, ``Liquid crystal hyperbolic metamaterial for wide-angle negative-positive refraction and reflection,'' Opt. Lett. {\bf 39}(7), 1744--1747 (2014).
\bibitem{27} V. I. Klyatskin, ``The imbedding method in statistical
boundary-value wave problems,''
Prog. Opt. {\bf 33}, 1--127 (1994).
\bibitem{28} K. Kim, H. Lim, and D.-H. Lee,
``Invariant imbedding equations for electromagnetic waves in stratified
magnetic media: Applications to one-dimensional photonic crystals,'' J. Korean Phys. Soc. {\bf 39}(6), L956--L960 (2001).
\bibitem{29} K. Kim, D.-H. Lee, and H. Lim, ``Theory of the propagation of
coupled waves in arbitrarily inhomogeneous stratified media,'' Europhys. Lett. {\bf
69}(2), 207--213 (2005).
\bibitem{30} K. Kim, D. K. Phung, F. Rotermund, and H. Lim, ``Propagation of electromagnetic waves
in stratified media with nonlinearity in both dielectric and
magnetic responses,'' Opt. Express {\bf 16}(2), 1150--1164 (2008).
\bibitem{31} S. Kim and K. Kim, ``Invariant imbedding theory of wave
propagation in arbitrarily inhomogeneous
stratified bi-isotropic media,'' J. Opt. {\bfseries 18}(6), 065605 (2016).
\bibitem{32} R. J. Potton, ``Reciprocity in optics,'' Rep. Prog. Phys. {\bf 67}(5), 717--754 (2004).

\end{thebibliography}
\end{document}